# Cross-Device Transfer of a Hospital-Trained Fetal Head Circumference Model to a Low-Cost Handheld Ultrasound Probe

Patrick Gao, Lara Hart, and Ken Sullivan

## Abstract

Background. Automated fetal head circumference (HC) measurement has reached near-expert accuracy on the HC18 benchmark, but published models are trained and evaluated on hospital-grade equipment. Whether such a model transfers to the low-cost handheld probes that could expand obstetric access in low- and middle-income countries has not been quantified.

Methods. A lightweight convolutional segmentation network (CSM, 0.13M parameters) was trained on HC18 across five independent runs and applied without modification to handheld images from a $1,350 probe (SonoQue C3), acquired at the same visit as the reference hospital scan. We measured the cost of input preprocessing, built a leave-one-out data-efficiency curve for fine-tuning over five random training-set draws at each size, and compared the model against two references: the hospital record, and measurements made on the handheld image itself.

Results. In-domain HC18 accuracy was 2.34 ± 0.43 mm (Dice 0.969 ± 0.002). On handheld images the same model achieved 16.12 mm against the hospital record (5.2% of the cohort mean head circumference of 312 mm), a 6.9-fold degradation. Preprocessing mattered: direct resizing gave 49.45 mm against 16.12 mm for an aspect-preserving two-pass pipeline, accounting for about a third of the gap. Fine-tuning on two images significantly worsened accuracy; from four images upward, error declined with training-set size at 0.75 mm per doubling ($p = 0.0045$), but no training-set size tested significantly improved on the unmodified model, the best being 1.37 mm at sixteen images (95% CI -0.51 to +3.25, $p = 0.143$). A single calibration constant estimated leave-one-out reduced disagreement with sonographer calipers on the same frame from 8.12 mm to 3.49 mm ($p = 0.0053$) but did not improve agreement with the hospital record. Measured against a manual tracing of the same image rather than the hospital record, the unmodified model was already within 10.48 mm, and agreement with sonographer calipers on the same frame was 8.12 mm ($r = 0.987$) with the difference negative in 10 of 10 cases (-8.12 ± 3.93 mm).

Conclusions. Fine-tuning on locally acquired images followed a clear data-efficiency relationship but did not surpass the unmodified model within the twenty-seven images available, and fine-tuning on very few images was actively harmful. Little correctable model error remained.

The unmodified model agreed more closely with a human tracing of the handheld image than with a hospital record acquired on different equipment, and its disagreement with sonographer calipers on the same frame was a systematic offset rather than scatter. After calibration, agreement with sonographer calipers was 3.49 mm (1.1% of mean HC), with 95% limits of agreement of ±8.6 mm or -2.84% to +2.78%, within the ±3.0% reported by Sarris et al. (2012) for a single observer remeasuring the same images. Cross-device evaluations should therefore report the input preprocessing used and characterise the precision of their reference standard, since each accounted for roughly a third of the apparent degradation.

---

# 1. Introduction

Obstetric ultrasound is essential to modern prenatal care, allowing for accurate estimation of gestational age, identification of fetal abnormalities, and monitoring of fetal growth throughout pregnancy [1]. Accurate gestational age estimation is particularly important in clinical decision-making; it provides crucial information regarding the timing of interventions such as corticosteroid administration for fetal lung maturation, induction of labor, and preparation for preterm delivery [2]. However, despite its clinical importance, access to obstetric ultrasound is largely limited in low- and middle-income countries (LMICs), where the majority of maternal and perinatal deaths happen. The World Health Organization estimates that approximately 260,000 women died during and following pregnancy and childbirth in 2023 [10], with approximately 92% of these deaths occurring in low- and lower-middle-income countries [11]. Neonatal mortality makes up over four million deaths annually worldwide, mainly in sub-Saharan Africa and south Asia, where access to skilled healthcare providers and diagnostic equipment remains critically insufficient [3].

## 1.1 The potential of portable ultrasound

In recent years, there has been a new generation of low-cost, handheld portable ultrasound devices that connect wirelessly to smartphones or tablets and cost between $700 and $5,000, a fraction of the $50,000 to $200,000 cost of conventional hospital-grade machines [4]. Studies have shown that these portable devices, including the Butterfly iQ and Clarius C3, are highly accurate for manual fetal biometric measurements when operated by trained sonographers, with gestational age estimates within one to two days of estimates by hospital-grade equipment [5]. Thus, portable ultrasound has the potential to greatly expand access to obstetric care in resource-limited settings. However, a critical barrier remains: the global shortage of trained sonographers capable of acquiring and interpreting ultrasound images. In sub-Saharan Africa, access to skilled operators for obstetric ultrasound is as low as 40%, and less than 30% in south Asia, meaning that even where devices exist, the human expertise required to use them effectively is often absent [3].

## 1.2 Artificial intelligence as a solution

Artificial intelligence, particularly deep learning, is a promising approach to automating obstetric ultrasound interpretation, possibly allowing minimally trained healthcare workers to take clinically meaningful measurements on their own [9]. Convolutional neural networks have been used to successfully automate estimation of gestational age from fetal biometric parameters such as head circumference, biparietal diameter, abdominal circumference, and femur length [2]. Fetal head circumference is one of the most important parameters; it is used for accurate gestational age estimation across all trimesters and is relatively robust to fetal growth restriction, a condition prevalent in LMIC populations where maternal malnutrition is common [2]. Several deep learning models have performed at near-expert levels on automated fetal head circumference measurement, with mean absolute differences below 2 mm and Dice similarity coefficients exceeding 97% on the publicly available HC18 benchmark dataset [6,7,8].

## 1.3 The domain gap problem

Despite these advances, a problem remains that limits the real-world applicability of these models in resource-limited settings. The works cited above are all trained and evaluated on images from high-specification hospital ultrasound machines, and all report their accuracy on HC18. On portable devices, image quality differs substantially due to reduced processing power, lower transducer element density, and unfamiliar artifact patterns, which can result in lowered contrast and increased noise [2]. Model performance is known to degrade when a network is applied to images from equipment other than that used for training, a phenomenon termed domain shift, and this has been documented across medical imaging modalities [13]. A scoping review of AI applications for point-of-care ultrasound in low-resource settings identified generalizability across devices as one of the most critical barriers to real-world deployment [9].

The low-resource problem has been approached, but from the opposite direction. van den Heuvel et al. [14] - the group that produced HC18 - developed automated HC estimation from free-hand sweeps for use in resource-limited countries, collecting and annotating a new dataset on a Samsung SonoAce R3 and measuring the reference on the same device. That approach works, but it requires device-specific data collection and expert annotation for each new device, which is precisely the resource the setting lacks. The complementary question - whether an existing hospital-trained model can be carried across to a low-cost probe, and at what cost in accuracy - has not been quantified for fetal head circumference, nor has it been established how much local data domain adaptation would require.

## 1.4 The present study

We evaluate a lightweight convolutional network for automated fetal HC measurement, trained on hospital-grade HC18 images and applied to images acquired with a $1,350 handheld probe. We quantify the degradation; isolate the contribution of input preprocessing, which is rarely reported and which we show to be a threefold effect; construct a data-efficiency curve to establish how many locally acquired images fine-tuning requires; and compare the model

against measurements made on the handheld images themselves, separating model error from the variance of a reference acquired on different equipment.

---

# 2. Methods

## 2.1 HC18 dataset

This study utilized the HC18 dataset introduced by van den Heuvel et al. [12], a publicly available benchmark for automated fetal head circumference measurement. The dataset comprises 1,334 two-dimensional ultrasound images collected from 551 pregnant women at Radboud University Medical Center, Nijmegen, Netherlands, between May 2014 and May 2015. Images were taken by experienced sonographers using either the GE Voluson E8 or GE Voluson 730 ultrasound system at the standard transventricular plane used for HC measurement. Each image is 800 × 540 pixels with pixel sizes ranging from 0.052 to 0.326 mm, reflecting variation in depth settings and zoom levels across examinations. The dataset is divided into 999 training images with corresponding expert-annotated HC measurements and 335 test images without publicly available expert annotations. Gestational age ranged from approximately 13 to 40 weeks. For this study, the 999 training images were divided into training and validation sets at an 80:20 ratio following the preprocessing procedure described in Section 2.3, yielding approximately 800 images for training and 200 for validation.

## 2.2 Model architecture

We adopted the Convolutional Segmentation Machine (CSM), a lightweight deep convolutional neural network developed by Zeng et al. [7] for efficient automated fetal head segmentation from two-dimensional ultrasound images. The CSM architecture was designed for deployment in resource-constrained settings, containing only 0.13 million parameters - substantially fewer than architectures such as U-Net - while maintaining competitive segmentation accuracy. The model takes a two-dimensional ultrasound image as input and produces a binary segmentation mask where white pixels indicate the fetal head region and black pixels indicate background. This lightweight design makes CSM a practical choice for portable device deployment where computational resources are limited.

## 2.3 Preprocessing and training

All HC18 training images were cropped to a uniform resolution of 768 × 512 pixels. Data augmentation was applied to the training set to increase image diversity and reduce overfitting, comprising horizontal and vertical flipping, rotations of ±10°, ±20° and ±30°, and scaling by 0.85 and 1.15. Augmented images and their corresponding binary masks were written to disk prior to training. No augmentation was applied to the validation set.

The CSM model was trained from scratch with no pretrained weights, using the Adam optimizer with a learning rate of 0.001 and a batch size of 32 for 10 epochs, on an Apple M3 processor using the Metal Performance Shaders backend. Because model weights are randomly initialized, training was repeated across five independent runs; all results below are reported across these five checkpoints, and the resulting standard deviation is stated separately from patient-to-patient variability wherever both are relevant.

## 2.4 Post-processing and HC measurement

Following segmentation, the largest connected component was extracted from the binary mask to eliminate spurious predictions. Edge detection was applied to obtain the outline of the fetal head, and a least-squares ellipse fitting algorithm was fitted to the extracted contour, yielding centre coordinates, semi-major and semi-minor axis lengths, and rotation angle. Head circumference was calculated from the fitted ellipse semi-axes using the Ramanujan approximation. The measurement in pixels was converted to millimetres using the per-image pixel size, taken from the HC18 metadata for benchmark images and from the calibration procedure of Section 2.6 for portable images.

The predicted mask was mapped back through the exact inverse of its own preprocessing transform, and the ellipse fitted at the image's native resolution, before conversion to millimetres. Comparisons between preprocessing strategies therefore isolate preprocessing alone.

## 2.5 Evaluation metrics

Mean difference (MD) is the signed average difference between predicted and reference HC in millimetres, indicating systematic over- or underestimation. Mean absolute difference (MAD) is the average magnitude of that difference regardless of direction, and is the primary accuracy metric. Dice similarity coefficient (DSC) measures overlap between predicted and reference masks, calculated as $2|A \cap B| / (|A| + |B|)$. Hausdorff distance (HD) is the maximum distance between predicted and reference boundaries in millimetres. DSC and HD are reported for HC18, where pixel-level expert annotations are available.

Agreement between measurement methods is additionally reported as bias with 95% limits of agreement, following Bland-Altman convention, because mean absolute difference alone does not distinguish a systematic offset from random scatter. The distinction matters because an offset can be removed by a single calibration constant where scatter of the same magnitude cannot, and the model's disagreement proves to be largely offset against same-frame human measurements and largely scatter against the hospital record (Sections 3.4 and 4.3). Mean absolute difference is averaged across the five checkpoints, so that its standard deviation describes run-to-run stability. Agreement statistics require a single measurement per patient and are therefore computed on the checkpoint-averaged prediction; this yields a slightly different mean absolute difference for the same comparison, and figures from the two conventions are not interchangeable. Bias is unaffected by the choice, since averaging is linear and the mean signed difference is the same under either convention.

## 2.6 Portable device dataset

Portable images were acquired with a SonoQue C3 wireless handheld convex probe (approximate retail cost $1,350) at a single obstetric clinic in Georgia, USA between July and October 2026, at the transventricular plane following the same anatomical criteria as HC18. Native frames are 987 × 1424 pixels after removal of device interface elements, an aspect ratio of 0.69 against the 1.50 expected by the network. The practice has no affiliated institutional review board. Verbal informed consent was obtained from each patient before image acquisition, and images were de-identified before analysis; see the Ethics Declaration.

Each portable image is paired with a head circumference measured at the patient's routine hospital examination on hospital-grade equipment. The handheld and hospital scans were performed on the same day by the same sonographer, so the two measurements describe the fetus at the same gestational age. The hospital measurement is the reference standard throughout.

Pixel spacing was determined per image by detecting the depth ruler graduations in the device overlay and computing millimetres per pixel from the known 5 mm tick interval, deriving calibration from the image itself rather than from recorded settings.

Of 32 acquisitions, four lacked a usable image or a reference measurement. Each of the remaining 28 images was manually annotated by tracing the outer skull boundary, and HC computed from that tracing by the same post-processing pipeline used for model predictions. Ten of the 28 had a tracing differing from the hospital record by more than 10% of the recorded value, indicating that the record did not correspond to the image; one such case is annotated as mislabelled at acquisition in the source log. The disagreements were bimodal: seventeen cases at or below 5.2%, one at 8.0%, and ten at or above 17.0%. Any threshold between 9% and 16% therefore produces the same partition, so the 10% rule is not a tuned parameter. Eighteen cases were analysed. Because exclusion concerned the validity of the record rather than the quality of the image or its tracing, all 28 images were retained in the fine-tuning pool (Section 2.8). One pair of acquisitions is recorded as the same fetus entered under two patient numbers; only one contributes to any analysis.

## 2.7 Input preprocessing strategies

Because the portable aspect ratio differs substantially from the network input, the method used to reconcile them is a free parameter that is seldom reported. We evaluated three strategies on the same images with the same five checkpoints. Direct resize resamples to the network input dimensions without regard to aspect ratio, compressing the head by approximately 2.3× along one axis. Letterbox rescales isotropically to fit within the input and zero-pads the remainder, introducing no distortion but leaving the head occupying a reduced fraction of the available resolution. Two-pass runs a letterbox pass to localize the head, then crops an aspect-matched window around it at native resolution and runs a second inference pass; this is the pipeline used for all other results.

No corresponding comparison is reported on HC18, where images are 768 × 512 and the aspect ratio is exactly 1.50, making letterboxing and direct resizing the identical computation.

## 2.8 Fine-tuning protocol

Fine-tuning used leave-one-out cross-validation over the 18 analysed patients. For each held-out patient and each training set size n in {0, 2, 4, 8, 16, 27}, a fresh copy of the pretrained network was fine-tuned on n patients drawn at random from the remaining pool and evaluated on the held-out patient; n = 0 is the unmodified pretrained model. The training pool comprised all 28 annotated patients excluding the held-out one. This includes the ten whose hospital record did not correspond to their image: fine tuning is supervised by the manual tracing, which describes its own image correctly regardless of which record was attached to it, so those cases are valid training data. Since evaluation requires a valid hospital measurement, they were not used as held-out cases.

The first convolutional stage was frozen and the remainder updated with Adam at a learning rate of 0.001 for a fixed budget of 600 gradient steps at every n, so that the curve measures the effect of data quantity rather than optimisation budget. Batch size was min(8, n). Augmentation matched the HC18 training operations of Section 2.3, applied on the fly with a single transform shared by each image and its mask; the held-out image was never augmented. Five random draws were evaluated at each (patient, n) combination, giving 468 fine-tuning runs.

Every random operation within a run - the choice of training subset, the augmentation draws, and the order in which images were presented within each epoch - was seeded from that run's own identifier, so a run is reproducible independently of the order in which runs are executed. Disagreement between draws at the same n therefore reflects the sampled training subset and the stochasticity of optimisation, and nothing else. At n = 27 the pool excluding the held-out patient contains exactly 27 patients, so every draw trains on the same images and the spread between draws at that size measures optimisation stochasticity alone.

The learning rate was verified rather than assumed. At n = 27, where the training subset is fixed, rates of 0.001, 0.0001 and 0.00001 were compared on nine patients at three draws each. Lower rates reduced the spread between draws but left a higher final training loss and worse accuracy, indicating undertraining rather than improved convergence; 0.001 gave both the lowest training loss and the lowest error and was retained.

Fine-tuning used the single-pass letterbox pipeline rather than the two-pass pipeline, and improvements are reported relative to that arm's own n = 0 baseline, which is 17.06 mm against the hospital record and 9.49 mm against the manual tracings. The two-pass figures of Section 3.1 are therefore not directly comparable to the curve.

## 2.9 Same-image reference comparisons

The hospital record measures the same fetus on different equipment, so disagreement with it contains both model error and the variance of that comparison. Two measurements of the handheld image itself allow the two to be separated. The manual tracings of Section 2.6 provide one. For ten patients a head circumference measured by the sonographer at acquisition was also available from the device overlay, with the ellipse placed manually and the perimeter computed by the device; the device's automatic biometry function was not used. Both references are reported throughout. The hospital record is the clinically relevant comparison, since it is what the care pathway would record and act on; measurements made on the handheld image itself isolate model error from the variance of a cross-equipment comparison.

These two references answer different questions. Measurements made on the handheld image itself ask whether the model can replace the person measuring at the point of care, which is the comparison that matters for the low-cost deployment this work targets. The hospital record asks whether a handheld device and an automated measurement can reproduce what a referral centre would record, which is the standard the pathway would eventually have to meet. Both are reported throughout and they are not interchangeable.

## 2.10 Statistical analysis

Paired comparisons used the Wilcoxon signed-rank test, which does not assume normality of differences and is appropriate given the sample size; the paired t test is reported alongside it where both were computed. Correlations are Pearson's r. Agreement between measurement methods is reported as bias, the mean signed difference, with 95% limits of agreement given by the bias plus or minus 1.96 standard deviations of the differences, expressed both in millimetres and as percentages of the mean of each measurement pair. Interpreting these requires an acceptance criterion. The ISUOG biometry guideline [16] specifies the mechanism but not the value, directing centres to audit a sample of stored images for inter- and intraobserver reproducibility and to retrain operators whose measurements fall persistently outside the 95% limits of agreement. The values are taken from Sarris et al., who measured 175 scans of 140 fetuses between 14 and 41 weeks with three experienced sonographers and report 95% limits of agreement for head circumference by the ellipse method of ±3.0% intraobserver and ±4.9% interobserver. Interobserver agreement is the primary criterion here, as every comparison reported crosses observers. The comparison is made in percentages rather than millimetres because the reference cohort spans 14 to 41 weeks whereas the present cohort is third trimester with a mean head circumference of 311.7 mm, so equal relative agreement corresponds to a larger absolute figure here. Limits of agreement assume approximately normal differences, which not all comparisons below satisfy; the proportion of cases within each reference band is therefore reported alongside, as it carries no distributional assumption.

Comparison of each fine-tuning set size against the unmodified model is the primary analysis of the curve. Two secondary analyses are also reported and were specified after the primary results were seen, which we state explicitly because it bears on how they should be weighted. The first pools all set sizes of four or more into a single paired comparison against the unmodified model. The second tests for a trend: for each patient, the slope of error against log2(n) was fitted across the fine-tuned conditions and the resulting slopes tested against zero in a one-sample test. A dose-response design carries more power in a trend test than in separate pairwise comparisons because it uses the ordering of the conditions, but the trend test does not include the unadapted condition and therefore speaks only to the effect of training-set size among fine-tuned models. It is reported both across all fine-tuned sizes and across sizes of four or more, since the n = 2 condition is worse than no adaptation and violates the assumed monotonicity.

A constant-calibration analysis is also reported. For each patient an additive constant, the mean of reference minus prediction, and a multiplicative factor, the mean of reference divided by prediction, were estimated from the remaining patients and applied to the held-out one; the resulting absolute error was compared with the uncalibrated value by paired test. No patient contributed to its own correction. This analysis is post-hoc, having been prompted by the pattern of results in Section 3.4.

Analysis used Python 3.8.20, PyTorch 2.4.1, NumPy 1.24.4, SciPy 1.10.1, pandas 2.0.3, OpenCV 4.13.0 and scikit-learn 1.3.2.

---

# 3. Results

## 3.1 In-domain and cross-device accuracy

Across five independent training runs evaluated on 200 held-out HC18 validation images, the CSM model achieved a mean absolute difference of 2.34 ± 0.43 mm, Dice similarity coefficient of 0.969 ± 0.002, and Hausdorff distance of 0.93 ± 0.06 mm. This is consistent with performance reported for this architecture [7] and confirms that the training pipeline reproduces published in-domain accuracy. Applied without modification to the 18 handheld images, the same five checkpoints achieved a mean absolute difference of 16.12 mm, a 6.9-fold increase (Table 1).

| **Table 1. Measurement accuracy in-domain and across devices. Values are mean ± SD across five independently trained models; the two-pass pipeline is used throughout.** | | | | | | | |
|---|---|---|---|---|---|---|---|
| | | | | | | | |
| **Dataset** | n | MAD (mm) | Bias (mm) | SD across patients (mm) | Dice | Hausdorff (mm) | vs HC18 |
| **HC18 (in-domain)** | 200 | 2.34 ± 0.43 | - | - | 0.969 ± 0.002 | 0.93 ± 0.06 | 1.0x |
| **Handheld, verified** | 18 | 16.12 ± 0.62 | -3.45 | 13.32 | - | - | 6.9x |

**Table 1.** Measurement accuracy in-domain and across devices. Values are mean ± SD across five independently trained models; the two-pass pipeline is used throughout. Dice and Hausdorff require pixel-level annotations and are defined only for HC18.

Two sources of variability are distinguished. Across the five checkpoints the standard deviation was 0.62 mm, indicating the result is not an artifact of any single training run. Across patients the standard deviation was 13.32 mm, which is the clinically relevant spread and reflects substantial case-to-case variation in how well the model transfers.

Accuracy under each preprocessing strategy is shown in Figure 1, and agreement between the model and the hospital record in Figure 2.

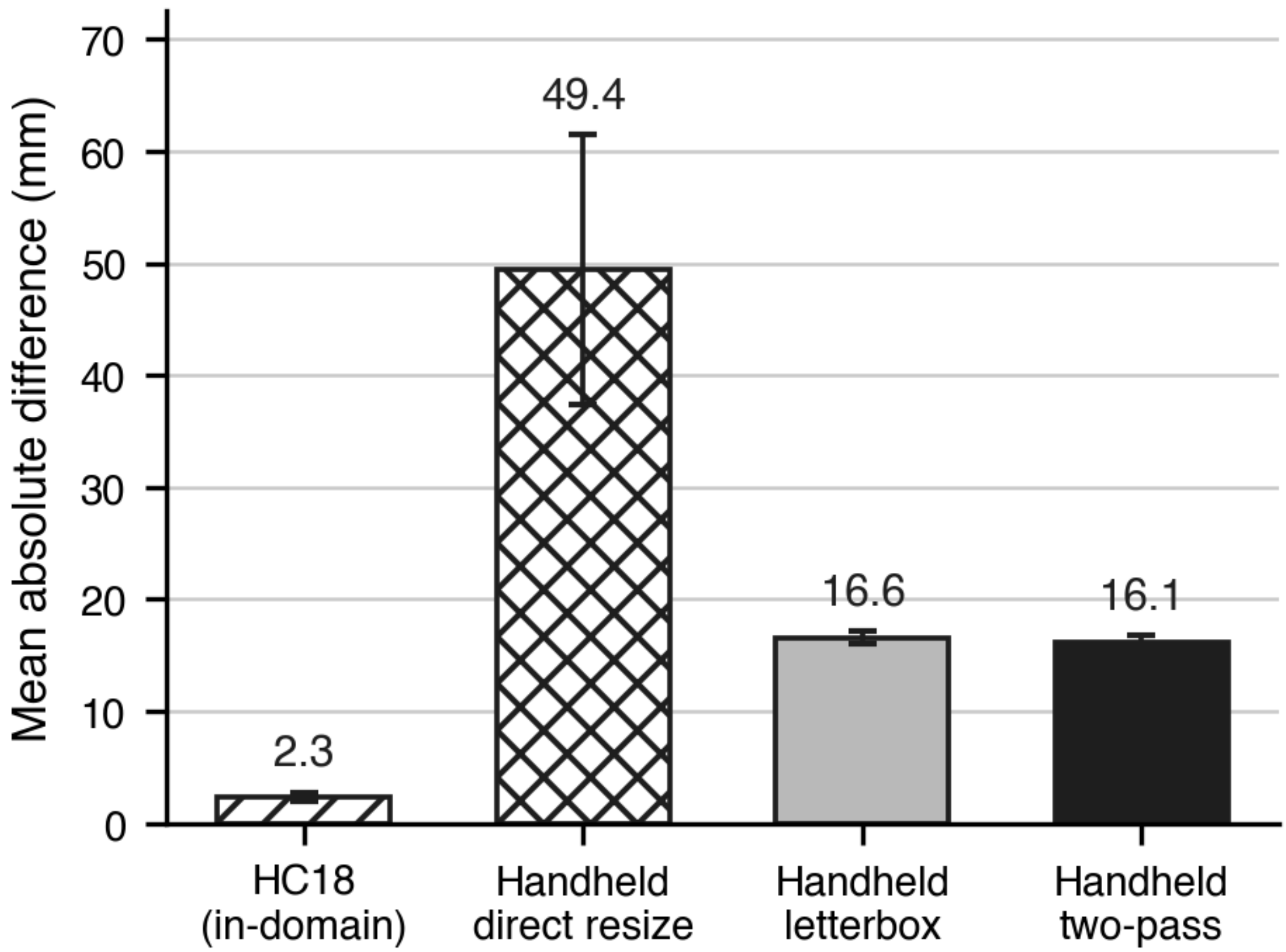

**Figure 1.** Mean absolute difference between automated and reference head circumference. HC18 is in-domain validation (n = 200, five independently trained models). Handheld bars are the same five models applied to the 18 analysed portable images under three input preprocessing strategies. Error bars are standard deviation across the five models.

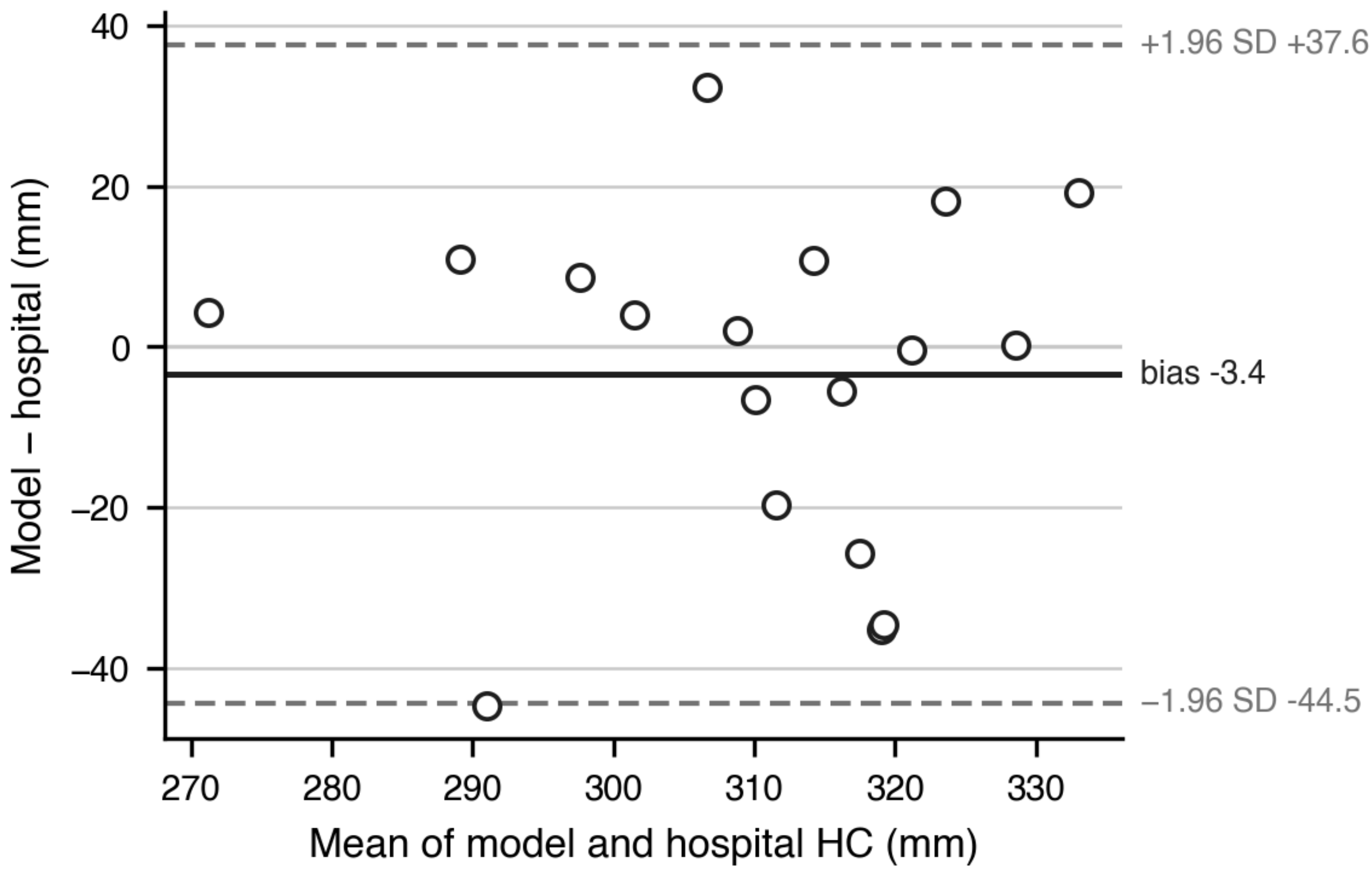

**Figure 2.** Bland-Altman plot of model against hospital record for the 18 analysed portable cases, using the two-pass pipeline averaged over five models. Solid line, bias; dashed lines, 95% limits of agreement.

### 3.2 Contribution of input preprocessing

Preprocessing strategy had a substantial effect on apparent accuracy. Direct resizing produced a mean absolute difference of 49.45 mm, against 16.59 mm for letterboxing and 16.12 mm for the two-pass pipeline, evaluated on the same images with the same checkpoints. Naive preprocessing therefore inflates the apparent device gap approximately threefold.

The direct-resize condition fails in two further ways. Its bias equals its mean absolute difference, -49.45 mm, meaning it under-measured in every case: a head compressed along one axis fits an ellipse that is too small. Additionally, its standard deviation across the five models was 12.03 mm, against 0.57 mm for letterboxing and 0.62 mm for the two-pass pipeline, so it is markedly less stable between training runs as well as less accurate.

Preprocessing does not explain the degradation. With the best available pipeline the gap remains 6.9-fold, so aspect-ratio distortion accounts for roughly one third of the total. The two-pass pipeline improves on letterboxing by 0.47 mm.

### 3.3 Fine-tuning data-efficiency curve

Without adaptation the letterbox arm gave 17.06 mm against the hospital record. Fine-tuning on two images made this significantly worse. Against the manual tracings, which are what the network was trained to reproduce, error rose from 9.49 mm to 11.79 mm, a deterioration of 2.31 mm (95% CI -3.66 to -0.95, p = 0.0022), and the proportion of cases in gross error by more than 40 mm rose from 5.6% to 10.0%. Under a fixed 600-step budget the network sees those two heads repeatedly and nothing else, fitting them at the expense of the general shape learned from 800 HC18 images.

From four images upward, accuracy improved steadily with training-set size (Table 2, Figure 3). Error against the record fell to 17.94 mm at four images, 16.93 mm at eight, 15.69 mm at sixteen and 16.00 mm at twenty-seven. Gross failures fell monotonically across the same range, from 7.8% to 3.3%, and the spread between training draws at a given size fell from 12.87 mm to 5.72 mm, so larger local datasets produce not only more accurate but more repeatable adaptation.

No individual training-set size improved significantly on the unmodified model. The largest improvement was 1.37 mm at sixteen images (95% CI -0.51 to +3.25, p = 0.143), and pooling all sizes of four or more gave 0.42 mm (95% CI -1.15 to +1.98, p = 0.579). With 18 patients and 13.32 mm of variation between them, the study is powered to detect a reduction of roughly 3 mm, and the observed effects are smaller than that.

The trend across set sizes is significant. Error declined by 0.747 mm per doubling of training-set size against the record (95% CI -1.229 to -0.265, p = 0.0045) and 0.771 mm per doubling against the tracings (95% CI -1.183 to -0.358, p = 0.0010), with the slope negative in 14 of 18 patients under both references. Excluding the harmful n = 2 condition leaves the relationship intact and, against the record, marginally steeper: -0.784 mm per doubling (95% CI -1.358 to -0.209, p = 0.0104) and, against the tracings, -0.620 mm per doubling (95% CI -1.147 to -0.093, p = 0.0237). The trend is therefore not an artifact of recovery from the two-image condition. It does not, however, establish that fine-tuning surpasses no adaptation, since it is fitted across the fine-tuned conditions only.

| **Table 2** | | | | | | | |
|---|---|---|---|---|---|---|---|
| | | | | | | | |
| **Training images (n)** | **MAD vs record (mm)** | **MAD vs tracing (mm)** | **Difference vs n = 0 (mm)** | **95% CI** | **p** | **Gross failures >40 mm** | **Between-draw spread (mm)** |
| **0** | 17.06 | 9.49 | - | - | - | 5.6% | - |
| **2** | 18.38 | 11.79 | -1.32 | -3.12 to +0.47 | 0.139 | 10.0% | 19.64 |
| **4** | 17.94 | 10.59 | -0.88 | -2.68 to +0.92 | 0.315 | 7.8% | 12.87 |
| **8** | 16.93 | 9.53 | 0.13 | -1.28 to +1.55 | 0.848 | 6.7% | 11.35 |
| **16** | 15.69 | 9.53 | 1.37 | -0.51 to +3.25 | 0.143 | 4.4% | 8.60 |
| **27** | 16.00 | 8.67 | 1.06 | -0.96 to +3.09 | 0.285 | 3.3% | 5.72 |

**Table 2.** Fine-tuning data-efficiency curve. Each row is the mean over 18 held-out patients and five training-set draws, at a fixed budget of 600 gradient steps. Differences are against the n = 0 condition in the same arm, positive where fine-tuning reduced error, with paired t test p values. Between-draw spread is the mean within-patient range of predicted HC across the five draws; at n = 27 every draw trains on the same images, so it reflects optimization stochasticity alone.

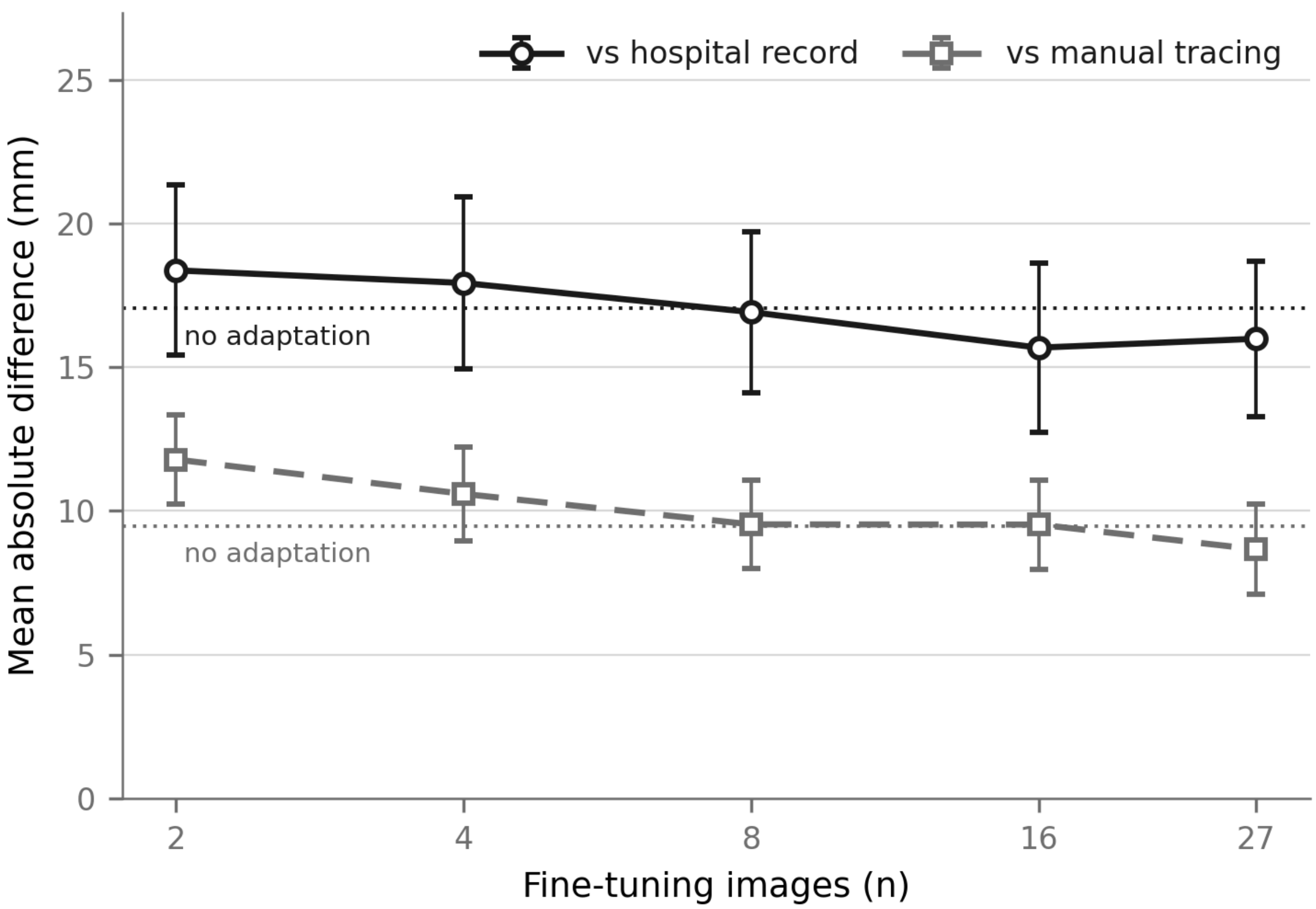

**Figure 3.** Data-efficiency curve for fine-tuning, scored against both references. Each point is the mean absolute difference over 18 held-out patients and five training-set draws at a given training set size, with a fixed budget of 600 gradient steps at every n. Error bars are standard error across patients. Dotted lines mark performance without adaptation for each reference.

How much room fine-tuning had is set by the annotations it trains on. The unmodified model was already within 9.49 mm of the tracings, and the tracings themselves differ from the hospital record by 10.44 mm, so a model that reproduced the tracings perfectly would still score about 10.44 mm against the record rather than zero. The improvement available against the record is therefore about 6.6 mm, of which the best fine-tuned condition realised roughly a fifth.

## 3.4 Agreement between measurement methods

Across the 18 analysed cases, manual tracings of the portable images and the hospital records differed by 10.44 mm on average, 3.34% of HC, with bias +1.94 mm and 95% limits of agreement of -20.8 to +24.7 mm, or -6.63% to +7.95%. Sarris et al. [15] report 95% limits of agreement for HC of ±3.0% within an observer and ±4.9% between observers measuring the same images under a standardised protocol. Our comparison is wider than either, which is expected, since it adds equipment and session variance to observer variance. It is not, however, uniformly wider: 16 of the 18 tracings fall within the ±4.9% interobserver band, and the limits of agreement are set by the two that do not. The differences are not normally distributed (Shapiro-Wilk p = 0.0087), so the proportion within the band characterises this comparison better than its limits do.

Scored against the manual tracings rather than the hospital record, the same five checkpoints gave 10.48 mm instead of 16.12 mm, a difference that is significant across the 18 patients (p = 0.016). Patient-to-patient spread also halved, from 13.69 mm to 7.15 mm.

In the ten patients with a sonographer-placed caliper measurement on the same frame, agreement separated by whether the comparison was made within or across equipment (Table 3).

| **Table 3. Agreement between measurement methods in the 10 patients with a sonographer-placed caliper measurement on the analysed frame.** | | | | |
|---|---|---|---|---|
| | | | | |
| **Comparison** | Frame | MAD (mm) | Bias (mm) | r |
| **Model vs on-device calipers** | Same | 8.12 | -8.12 | 0.987 |
| **Model vs manual tracing** | Same | 8.07 | -4.28 | 0.926 |
| **On-device calipers vs manual tracing** | Same | 5.94 | 3.84 | 0.955 |
| **Model vs hospital record** | Different | 12.69 | -1.40 | 0.595 |
| **On-device calipers vs hospital record** | Different | 14.38 | 6.72 | 0.696 |
| **Manual tracing vs hospital record** | Different | 9.80 | 2.88 | 0.798 |

**Table 3.** Agreement between measurement methods in the ten patients with a sonographer-placed caliper measurement on the analysed frame.

The three measurements of the same picture - the model, the sonographer's calipers and the manual tracing - agree with one another between 0.93 and 0.99, while every comparison against the hospital record falls to between 0.60 and 0.80. The sonographer's own on-device measurement differs from the hospital record by 14.38 mm, more than the model does. The three same-frame measurements agreed more closely with one another than any agreed with the hospital record, consistent with part of the disagreement arising from the reference rather than the model.

The model-on-device difference was negative in all ten cases, with a mean of -8.12 mm and a standard deviation of only 3.93 mm, so the model sits inside the sonographer's calipers by a near-constant amount rather than scattering around them. Across these ten cases the difference between same-image and cross-equipment agreement does not reach significance ($p = 0.43$). Both comparisons are shown in Figure 4, and the corresponding boundaries in Figure 5.

Because that difference is almost entirely systematic, it is correctable. A single additive constant estimated from the other patients and applied to the held-out one reduced disagreement with the sonographer's calipers from 8.12 mm to 3.49 mm (reduction 4.62 mm, 95% CI 1.76 to 7.48, $p = 0.0053$, improved in 8 of 10 patients); a multiplicative factor performed equivalently at 3.67 mm. The calibrated differences had 95% limits of agreement of ±8.6 mm, or -2.84% to +2.78%, within the ±3.0% reported by Sarris et al. [15] for a single observer remeasuring the same images. The same correction did not improve agreement with the hospital record and made it worse, from 15.74 mm to 17.43 mm ($p = 0.043$, improved in 7 of 18 patients), because the bias in that comparison is only -3.45 mm within a 20.93 mm standard deviation of differences and the estimated constant is pulled by the two largest outliers.

Table 4 expresses each comparison against the reproducibility criterion of Section 2.10. Only the calibrated same-frame measurement falls within the intraobserver band, and no uncalibrated comparison falls within the interobserver band, although the proportion of individual cases inside that band ranges from 56% against the hospital record to 90% against the sonographer’s calipers.

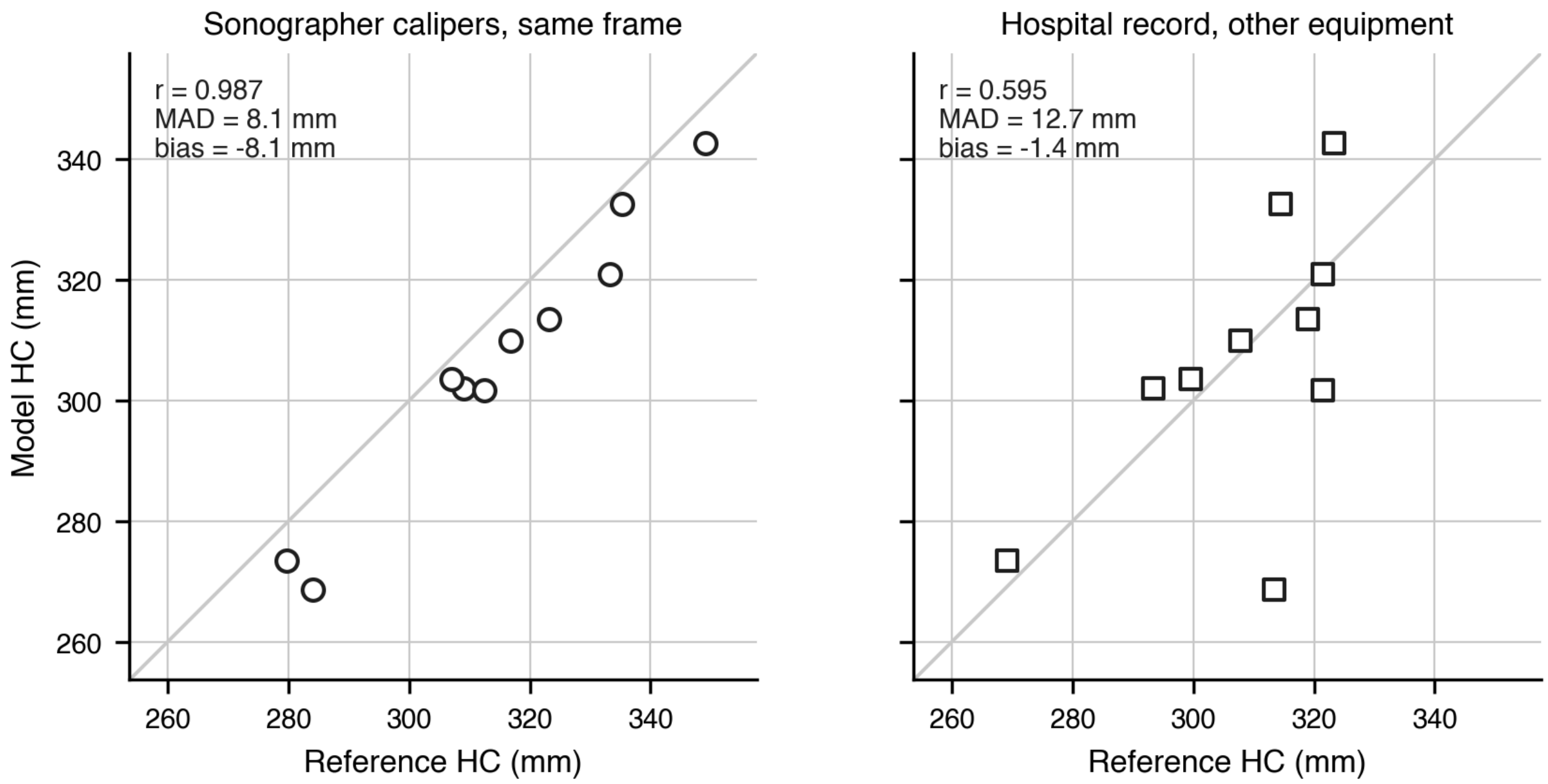

**Figure 4.** Model head circumference against two reference standards in the ten patients with a same-frame sonographer measurement. Left, calipers placed on the frame the model analysed; right, the hospital record acquired on different equipment. Diagonal is the line of identity.

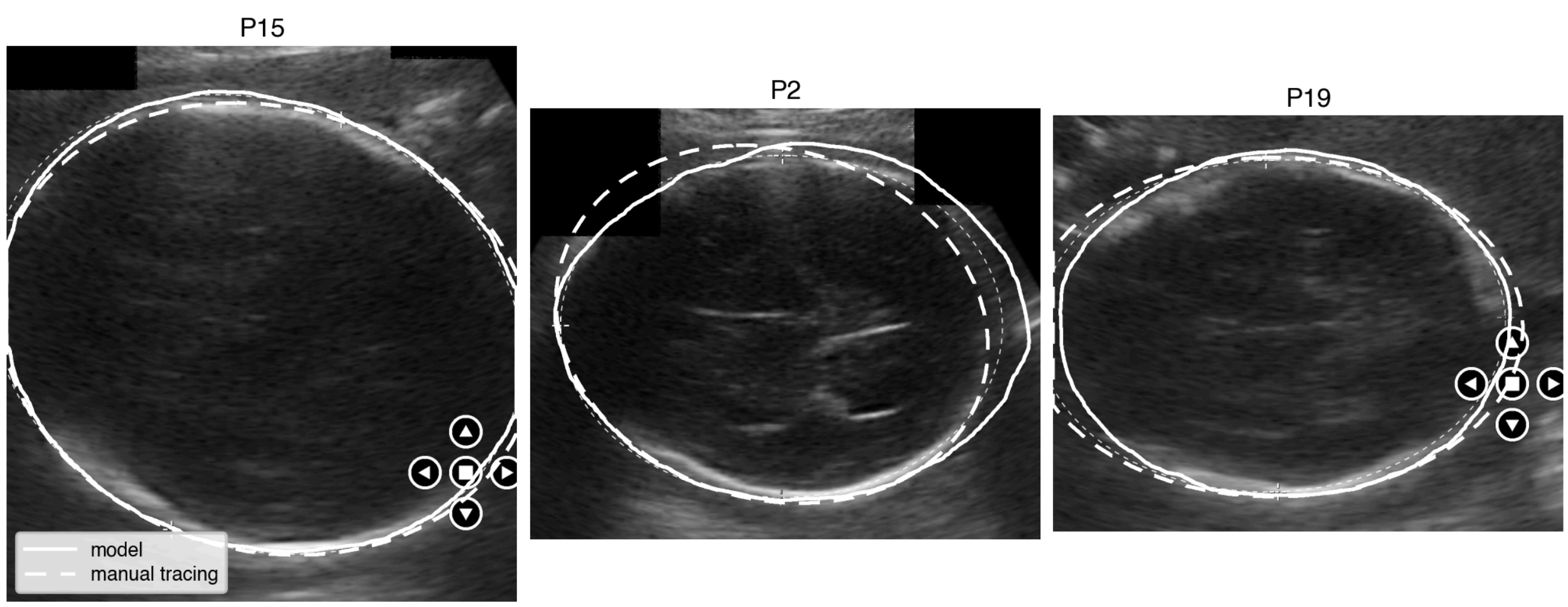

**Figure 5.** Boundary placement on three cases, selected as the minimum, median and maximum model error among patients with an available caliper image. Each panel is a single acquisition cropped to the fetal head, showing the sonographer's caliper ellipse (fine dotted curve, placed at acquisition), the model contour (solid) and the manual tracing (dashed). Across the ten cases the model contour lies inside the tracing, which lies inside the caliper ellipse. Measurements for these cases are given in Table 3.

**Table 4. Agreement against published reproducibility limits. Bias and 95% limits of agreement are computed on percentage differences, following Sarris et al. [15]. Reference bands are +/-4.9% interobserver and +/-3.0% intraobserver.**

| Comparison | n | Bias (%) | 95% LoA (%) | Within ±4.9% | Within ±3.0% |
|---|---|---|---|---|---|
| Model vs hospital record | 18 | -1.11 | -14.41 to +12.18 | 10/18 (56%) | 8/18 (44%) |
| Manual tracing vs hospital record | 18 | +0.66 | -6.63 to +7.95 | 16/18 (89%) | 9/18 (50%) |
| Model vs manual tracing | 18 | -1.77 | -8.75 to +5.20 | 13/18 (72%) | 10/18 (56%) |
| Model vs on-device calipers | 10 | -2.65 | -5.37 to +0.07 | 9/10 (90%) | 6/10 (60%) |
| Model + constant vs calipers | 10 | -0.03 | -2.84 to +2.78 | 10/10 (100%) | 10/10 (100%) |

**Table 4.** Agreement against published reproducibility limits. Bias and 95% limits of agreement are computed on percentage differences following Sarris et al. [15], and are symmetric about the bias rather than about zero. Reference bands are the ±4.9% interobserver and ±3.0% intraobserver limits reported in that study. Record comparisons use all 18 analysed patients; caliper comparisons use the 10 with a same-frame sonographer measurement. The constant in the final row was estimated leave-one-out. Differences for the tracing-versus-record comparison are not normally distributed (Shapiro-Wilk $p = 0.0087$), so its within-band proportion describes it better than its limits do.

---

# 4. Discussion

## 4.1 Principal findings

A hospital-trained model applied unmodified to a $1,350 handheld probe lost roughly a factor of seven in accuracy against the clinical record. Fine-tuning on locally acquired images followed a clear data-efficiency relationship but did not recover a statistically detectable amount within the twenty-seven images available, and on two images it was actively harmful.

## 4.2 The degradation is not a preprocessing artifact

Portable frames are portrait-oriented while the network expects a landscape input, so a naive implementation compresses the fetal head by roughly a factor of 2.3 along one axis. Any cross-device evaluation must resolve this mismatch somehow, and the choice is rarely stated in published work, so we measured its cost directly: a study using direct resizing would report approximately three times the device gap we measure with an aspect-preserving pipeline.

Preprocessing nevertheless accounts for only about a third of the degradation. With the best available pipeline the gap remains 6.9-fold, so the cross-device difficulty is real and is not explained by image geometry alone.

### 4.3 Error decomposes into a systematic offset and reference-standard variance

The hospital record measures the same fetus on different equipment. It is a sound clinical measurement, but it is a measurement rather than a known quantity, so any disagreement with it contains both the model's error and its own.

The two references therefore set different bars. Against measurements made on the image itself, the question is whether the model can stand in for the operator at the point of care. Against the hospital record, it is whether the whole low-cost pathway can reproduce a referral centre's number - a standard this study does not meet, and one that neither adaptation tested here moves toward, which locates the remaining work in acquisition and reference protocol rather than in the model.

Three measurements of the handheld image were available: the model's, the sonographer's calipers and a manual tracing. These agree with one another far more closely than any of them agrees with the hospital record, and the sonographer's own measurement differs from that record by more than the model does. Scored against the tracings rather than the record, the model's error falls from 16.12 to 10.48 mm and the spread between patients halves. Both point to the same conclusion that a substantial part of what the headline figure attributes to the model belongs to the comparison.

The scale of that comparison can be quantified. Sarris et al. [15] report 95% limits of agreement for HC of ±4.9% between observers measuring the same images under a standardised protocol. A manual tracing of the handheld image differs from the hospital record by -6.63% to +7.95%, wider than that band, as expected for a comparison that adds equipment and session differences to observer variance, and that width bounds what any model scored against this record can achieve. It does not account for the whole of the model's disagreement. The model's own limits against the record are roughly twice as wide again, -14.41% to +12.18%, and the model falls inside the interobserver band in 10 of 18 cases against 16 of 18 for the tracing. Part of the gap belongs to the reference and part to the model, and on these images the model is the less accurate of the two measurements made on them.

The systematic component has a practical consequence, and it is testable with the data in hand. An offset of 8.12 mm with a standard deviation of 3.93 mm, negative in every case, indicates the model's boundary sits consistently inside the sonographer's rather than varying around it. Both are marking the outer skull table; the difference is in where within that few-pixel band each places the edge. A single constant estimated from other patients removes most of it, bringing agreement with the sonographer's own calipers to 3.49 mm, with 95% limits of agreement of -2.84% to +2.78% - within the ±3.0% a single observer achieves remeasuring the same images [15].

That correction does not improve agreement with the hospital record; applied there it makes the error worse. Taken together the two results locate the offset precisely. It belongs to the comparison between the model and a human measuring the same frame, and not to the comparison between the model and a record acquired on other equipment, where the disagreement is scatter rather than offset. A clinic could calibrate this model to its own sonographer and reach agreement within human repeatability, and doing so would still not bring its output closer to what the referral hospital reports.

## 4.4 Fine-tuning follows a data-efficiency curve but had little room to improve

Three features of the curve are worth separating.

Fine-tuning on very few images is harmful, not merely unhelpful. At two images the model became significantly worse at reproducing the boundaries it had just been shown, and its rate of gross failure nearly doubled. The practical consequence is direct: a clinic cannot adapt a model from a handful of local scans, and attempting it would degrade the measurements it was meant to improve.

Beyond that point, more local data helps in a measurable and orderly way. Error falls by roughly 0.75 mm for each doubling of the local dataset, gross failures fall monotonically, and the variability between different draws of the same number of images falls by more than half. The last of these matters for deployment as much as accuracy does, because it determines whether a clinic's result depends on which particular scans it happened to collect.

But within the range this study could reach, the curve does not surpass the unadapted model. The reason is that little model error was available to remove. The unmodified model was already within 9.49 mm of a manual tracing of the same image, and tracings sit 10.44 mm from the record, leaving about 6.6 mm of recoverable error against the record; the best fine-tuned condition recovered roughly a fifth of it. The flat net result is therefore a statement about how little model error remained, combined with a sample too small to resolve what was recovered, rather than a failure of the method. With eighteen patients and 13.32 mm of variation between them, the study can detect a reduction of about 3 mm, and the effects observed are smaller.

The same conclusion is reached from the other direction by the calibration result of Section 3.4. Of the two adaptation strategies available without new expert annotation, neither gradient-based fine-tuning nor constant calibration improved agreement with the clinical record. Both are capable of removing model error, and calibration demonstrably does so against a same-frame human measurement. Neither improved agreement with the hospital record, a result consistent with that gap carrying a substantial reference-variance component alongside model error, since no adaptation of the model can reduce the reference’s own contribution.

## 4.5 Limitations

The reference standard bounds what this study can resolve. Acquiring both scans on the same day with the same sonographer removes growth and inter-observer variance, which is tighter than most validation studies manage, but the hospital measurement is made on different equipment and carries its own error of the order reported by Sarris et al. [15]. Separating residual model error from that variance more precisely would require expert measurements made on the handheld images themselves across the full cohort.

Of 28 annotated acquisitions, ten were excluded because the record did not correspond to the image. That is a limitation of the data collection rather than the analysis, and a cohort in which more than a third of records could not be reconciled with their images warrants stricter acquisition-time verification in future work.

Manual tracings were produced with the corresponding hospital HC visible to the annotator, which can bias a tracing toward the target value and would make tracing-record agreement appear better than a blind protocol would yield. The 10.44 mm figure should therefore be read as a lower bound.

The largest training-set size is bounded by the dataset. At n = 27 only one subset of the pool exists, so that condition carries no information about how performance depends on which images are collected, only on the stochasticity of optimization. The curve therefore characterises local dataset sizes up to sixteen images with sampling variability, and twenty-seven without it.

The sample is small, drawn from a single clinic and a single operator, and includes one pair of acquisitions from the same fetus. All imaging used one handheld device, so the findings describe that device rather than low-cost ultrasound in general. Manual annotations were produced by a single non-expert annotator; inter-annotator variability was not assessed.

## 4.6 Implications and future work

Three implications follow for deployment of automated biometry on low-cost ultrasound. Preprocessing must preserve aspect ratio, and the choice should be reported, since it is a threefold effect and is invisible in papers that do not state it. Device-specific fine-tuning should not be attempted on a handful of images, since below roughly eight it degrades the measurement it is meant to improve, while the orderly improvement beyond that point suggests local adaptation is worth pursuing at larger scale. Where the operator measuring on the device is the same person whose measurements the model is intended to replace, a single calibration constant is a cheaper and more stable intervention than fine-tuning, requiring no training, no annotation and a handful of paired measurements. Finally, evaluation against clinical records from other equipment conflates model error with reference variance; where the goal is to characterise a model rather than a care pathway, reference measurements should be made on the images under test.

Future work should obtain expert measurements on the handheld images for a full cohort, which would both remove the dominant source of uncertainty here and allow the offset to be estimated directly. Resolving the 1.37 mm improvement observed at sixteen images would require approximately sixty cases at conventional power, and a 2 mm effect approximately thirty. A prospective test of calibration against expert measurements made on the handheld images themselves would establish whether the correction generalises beyond the cohort that suggested it.

A further direction would remove the annotation bottleneck rather than working within it. The ceiling identified here arises because the network is supervised by manual tracings, which are expensive to produce and carry their own variability. Supervising the measurement directly - propagating gradients from the fitted ellipse perimeter rather than from pixel overlap, for instance by deriving the ellipse axes from the spatial moments of the predicted mask, which is differentiable - would allow scalar clinical measurements already recorded in routine care to train the model without any new annotation. One scalar per image is far weaker supervision than a mask, so such an approach would need substantially more data than was available here, but it would address the constraint that makes device-specific adaptation impractical in exactly the settings this work targets.

---

# 5. Conclusions

Fine-tuning a hospital-trained fetal head circumference model on locally acquired handheld images produced a clear data-efficiency relationship - error fell by about 0.75 mm for every doubling of the local dataset - but no training-set size up to twenty-seven images significantly improved on the unmodified model, and fine-tuning on two images made it significantly worse.

The reason is that little correctable model error remained. Applied to a $1,350 handheld probe the unmodified model degraded 6.9-fold against the clinical record, but about a third of that is attributable to input preprocessing and is avoidable, and measured against a human tracing of the same image rather than a record from other equipment the model was already within 10.48 mm. Its disagreement with sonographer calipers on the same frame was a near-constant 8.12 mm offset, and a single calibration constant reduced that to 3.49 mm, comparable to the repeatability of a single human observer - yet the same correction did not improve agreement with the hospital record. Cross-device evaluations should therefore report the input preprocessing used and characterise the precision of their reference standard, and clinics deploying such a model should not expect to close the gap to a referral hospital's numbers by adapting it to local data.

---

# Ethics Declaration

This project evaluated the technical feasibility and performance of a novel ultrasound probe and AI algorithm. Because the protocol involved non-invasive imaging that did not alter clinical management or patient care, a formal institutional review board (IRB) review was not required. Nonetheless, we maintained rigorous ethical safeguards: we obtained verbal informed consent from all participants before image acquisition, and we fully de-identified all data before storage and analysis. This study was strictly utilized for non-interventional device evaluation and technical validation.

---

## Funding

This research received no specific grant from any funding agency in the public, commercial, or not-for-profit sectors. The handheld ultrasound probe was purchased by the authors.

---

## Competing Interests

The authors declare no competing interests. Neither author has any financial or consulting relationship with any ultrasound device manufacturer.

---

## AI Disclosure

The authors used Claude (Anthropic) during data analysis. All experimental design decisions, data collection, and interpretation of results were performed by the authors, who take full responsibility for the content.

---

## Data and code availability

Analysis code and derived results are available from the corresponding author on reasonable request. The HC18 dataset is publicly available [12]. Portable device images contain patient information and are not publicly released.

---

# References

[1] Self A, Chen Q, Desiraju BK, Dhariwal S, Gleed AD, Mishra D, Thiruvengadam R, Chandramohan V, Craik R, Wilden E, Khurana A; CALOPUS Study Group; Bhatnagar S, Papageorghiou AT, Noble JA. Developing Clinical Artificial Intelligence for Obstetric Ultrasound to Improve Access in Underserved Regions: Protocol for a Computer-Assisted Low-Cost Point-of-Care UltraSound (CALOPUS) Study. JMIR Res Protoc. 2022;11(9):e37374. doi:10.2196/37374

[2] Maraci MA, Yaqub M, Craik R, Beriwal S, Self A, von Dadelszen P, Papageorghiou A, Noble JA. Toward point-of-care ultrasound estimation of fetal gestational age from the trans-cerebellar diameter using CNN-based ultrasound image analysis. J Med Imaging (Bellingham). 2020;7(1):014501. doi:10.1117/1.JMI.7.1.014501

[3] Vasan A, Friend J. Medical Devices for Low- and Middle-Income Countries: A Review and Directions for Development. J Med Device. 2020;14(1):010803. doi:10.1115/1.4045910

[4] Ranger BJ, et al. Portable ultrasound devices for obstetric care in resource-constrained environments: mapping the landscape. Gates Open Res. 2024;7:133. doi:10.12688/gatesopenres.15088.2

[5] Rittenhouse KJ, Vwalika B, Sebastiao Y, Pokaprakarn T, Sindano N, Shah H, Stringer EM, Kasaro MP, Cole SR, Stringer JSA, Price JT. Accuracy of portable ultrasound machines for obstetric biometry. Ultrasound Obstet Gynecol. 2024;63(6):772-780. doi:10.1002/uog.27541

[6] Zeng Y, Tsui PH, Wu W, Zhou Z, Wu S. Fetal Ultrasound Image Segmentation for Automatic Head Circumference Biometry Using Deeply Supervised Attention-Gated V-Net. J Digit Imaging. 2021;34(1):134-148. doi:10.1007/s10278-020-00410-5

[7] Zeng W, Luo J, Cheng J, Lu Y. Efficient fetal ultrasound image segmentation for automatic head circumference measurement using a lightweight deep convolutional neural network. Med Phys. 2022;49(8):5081-5092. doi:10.1002/mp.15700

[8] Yang C, Yang Z, Liao S, Guo J, Yin S, Liu C, Kang Y. A new approach to automatic measure fetal head circumference in ultrasound images using convolutional neural networks. Comput Biol Med. 2022;147:105801. doi:10.1016/j.compbiomed.2022.105801

[9] Kim S, Fischetti C, Guy M, Hsu E, Fox J, Young SD. Artificial Intelligence (AI) Applications for Point of Care Ultrasound (POCUS) in Low-Resource Settings: A Scoping Review. Diagnostics (Basel). 2024;14(15):1669. doi:10.3390/diagnostics14151669

[10] World Health Organization. Maternal and newborn - Mortality/causes of death, 2021. https://www.who.int/data/maternal-newborn-child-adolescent-ageing/maternal-and-newborn-data/maternal-and-newborn---mortality-causes-of-death

[11] World Health Organization. Maternal mortality. Fact sheet. 7 April 2025. https://www.who.int/news-room/fact-sheets/detail/maternal-mortality. Accessed October 28, 2026.

[12] van den Heuvel TLA, de Bruijn D, de Korte CL, van Ginneken B. Automated measurement of fetal head circumference using 2D ultrasound images. PLoS One. 2018;13(8):e0200412. doi:10.1371/journal.pone.0200412

[13] Guo B, Lu D, Szumel G, Gui R, Li P, Wang T, Konz N, Mazurowski MA. The impact of scanner domain shift on deep learning performance in medical imaging: an experimental study. Int J Comput Assist Radiol Surg. 2026. doi:10.1007/s11548-026-03655-7

[14] van den Heuvel TLA, Petros H, Santini S, de Korte CL, van Ginneken B. Automated Fetal Head Detection and Circumference Estimation from Free-Hand Ultrasound Sweeps Using Deep Learning in Resource-Limited Countries. Ultrasound Med Biol. 2019;45(3):773-785. doi:10.1016/j.ultrasmedbio.2018.09.015

[15] Sarris I, Ioannou C, Chamberlain P, Ohuma E, Roseman F, Hoch L, Altman DG, Papageorghiou AT; International Fetal and Newborn Growth Consortium for the 21st Century (INTERGROWTH-21st). Intra- and interobserver variability in fetal ultrasound measurements. Ultrasound Obstet Gynecol. 2012;39(3):266-273. doi:10.1002/uog.10082

[16] Salomon LJ, Alfirevic Z, Da Silva Costa F, et al. ISUOG Practice Guidelines: ultrasound assessment of fetal biometry and growth. Ultrasound Obstet Gynecol. 2019;53(6):715-723. doi:10.1002/uog.20272